\documentclass[preprint,aps,nofootinbib]{revtex4}
\usepackage{}
\usepackage{graphicx}
\usepackage{amsmath}
\usepackage{amsfonts}
\usepackage{amssymb}
\usepackage{color}
\usepackage{dcolumn}
\providecommand{\U}[1]{\protect\rule{.1in}{.1in}}

\definecolor{lightgray}{rgb}{.7,.7,.7}

\definecolor{red}{rgb}{1,0,0}

\definecolor{blue}{rgb}{0,0,1}

\newcommand{\f}{\begin{equation}}
\newcommand{\ff}{\end{equation}}
\newcommand{\fa}{\begin{eqnarray}}
\newcommand{\ffa}{\end{eqnarray}}

\begin{document}
\title{Magnetohydrodynamics from gravity}
\author{Cheng-Yong Zhang $^{1}$}
\email{zhangchengy10@mails.gucas.ac.cn}
\author{Yi Ling $^{2}$}
\email{lingy@ihep.ac.cn}
\author{Chao Niu $^{1}$}
\email{niuchao09@mails.gucas.ac.cn}
\author{Yu Tian $^{1}$}
\email{ytian@gucas.ac.cn}
\author{Xiao-Ning Wu $^{3}$}
\email{wuxn@amss.ac.cn}

\affiliation{$^1$ College of Physical Sciences, Graduate University of Chinese Academy of Sciences, Beijing 100049, China\\ $^2$  Institute of High Energy Physics, Chinese
Academy of Sciences, Beijing 100049, China\\
$^3$ Institute of Mathematics, Academy of Mathematics and System
Science, CAS, Beijing 100190, China and Hua Loo-Keng Key
Laboratory of Mathematics, CAS, Beijing 100190, China\\}

\begin{abstract}
{Imposing the Petrov-like boundary
condition on the hypersurface at finite cutoff,} we derive the
hydrodynamic equation on the hypersurface from the bulk
Einstein equation with electromagnetic field {in the near
horizon limit}. We first get the general framework for spacetime
with matter field, and then derive the incompressible
Navier-Stokes equations for black holes with electric charge and
magnetic charge respectively. Especially, in the magnetic case, the standard magnetohydrodynamic equations will arise due to the existence of the background electromagnetic field on the hypersurface.
\end{abstract}

\maketitle

\section {Introduction}
Under the long wavelength limit, {the correspondence between
the bulk gravity and the fluid living on the horizon was firstly
discovered by Damour\cite{key-1}. He investigated the surface
effects in black hole physics and disclosed the similarity of the
behavior of the excitations of a black hole horizon to those of a
fluid.} After a series of studies, the fluid/gravity
correspondence was {extended to the boundary at infinity}
in the context of AdS/CFT
correspondence\cite{key-2,key-3,key-4,key-5,key-6,key-7,key-8,
key-9,key-10,key-11,key-12,key-13,MMC,JH,GC,CE}.
{Traditionally, this correspondence was constructed} by
directly disturbing the bulk metric under the regularity condition of
the horizon. It was
shown that in a {long wavelength} limit, {the
remaining dynamics on the boundary} is governed by the
incompressible Navier-Stokes equation{\cite{key-17,Cai,key-16,Cai2}}.

In \cite{key-18}, Lysov and Strominger firstly studied
the fluid/gravity correspondence by imposing the Petrov-like
condition on the {cutoff surface in the near horizon limit
and found that the constraint equations for gravity can
give rise to the incompressible Navier-Stokes equation for a fluid
living on a flat spacetime with less one dimensions. In this sense
the Petrov-like condition is of a holographic nature. Later,
this framework was generalized to {describe a dual fluid
living on a cutoff surface that is spatially curved in
\cite{key-19}. {In this case} the Navier-Stokes equation
{receives contributions from} the curvature of the
hypersurface. In \cite{key-20}, {the holographic nature of
the Petrov-like condition was further disclosed in a spacetime
with a cosmological constant}. {Based on all the work
above, it is reasonable to conjecture that in the approach of
fluid/gravity duality, imposing} the Petrov-like condition
{on the cutoff surface} is equivalent to {imposing}
the regularity {condition} on the future horizon at least
in the near horizon limit. However, technically, imposing Petrov-like condition is much simpler than imposing the regularity
{condition}.

In this paper, we will generalize this framework further to
a spacetime with matter fields. In
general, the presence of matter fields in a system involves more
degrees of freedom. Therefore, the key issue that we need to solve
is what kind of boundary condition should be imposed for matter
fields on the cutoff surface such that the remaining degrees of
freedom on the surface can lead to a right correspondence between
the} fluid and gravity. We will take the
electromagnetic field as an example and investigate how
to constrain its degrees of freedom so that the
remaining degree{s} of freedom of the
Einstein-Maxwell fields exactly matches
those of a charged fluid\cite{key-14,key-15}. Under this framework, we explicitly show
that the (magneto)hydrodynamics of the dual charged fluid is governed by the incompressible
Navier-Stokes equations, where an external force will arise if a
background electromagnetic field exists on the cutoff surface.

We organize the paper as follows. In section II, we
firstly construct the general framework for
imposing Petrov-like condition on the cutoff surface with matter
fields, and then propose a boundary condition for the
electromagnetic field on the cutoff in detail. In section III, we
apply the framework to the} charged $AdS_{p+2}$ black brane
and then derive the incompressible Navier-Stokes equation
on a flat hypersurface in the near horizon
limit. In section IV, we consider a spacetime with
Reissner-Nordstrom-AdS (RN-AdS) {background in} which
{the embedded hypersurface is intrinsically curved}. In section V, the magnetic black hole is considered.
{We find an external force will arise due to the
existence of the background electromagnetic field {on the hypersurface}.} Our main
results and conclusions are summarized in Section VI. In appendix
A, we present a detail{ed} calculation {on the
perturbations of the electromagnetic field}. Appendix B {is
on} the specific form{s} of {the} Hamiltonian
constraint {in the background of} charged AdS black brane
and RN-AdS spacetime, {respectively}.

\section{The general framework for \textbf{a} spacetime with matter field\textbf{s}}

We {construct the general framework for imposing the Petrov-like condition on the cutoff surface in the presence of matter.
Our general setup is following. Firstly, we require that the
dynamics of the $p+2$ dimensional bulk spacetime is subject to the
standard Einstein equation with matter. Secondly, the background
contains a Killing horizon. Then given an embedded
hypersurface which has a finite distance from the horizon, we
consider the perturbations of the extrinsic curvature as well as
the matter fields on the cutoff surface under appropriate boundary
conditions. In this section we will consider the boundary
condition for gravity at first, and then for Maxwell field
later.}

{For the gravity part such a boundary condition is proposed to
be the Petrov-like condition. It should be noticed that the
Petrov conditions in its original form are proposed to
classify the geometry of the whole spacetime such that it can be
defined at each point in the bulk spacetime. For our purpose we
only specify this condition on the cutoff surface, therefore the
first thing we need to do is decomposing the $p+2$ dimensional
Weyl tensor appearing in Petrov-like conditions in terms of the
$p+1$-dimensional quantities, which includes the intrinsic
curvature and extrinsic curvature of the hypersurface as well as
the induced metric on it.}

Consider a $p+2$ dimensional spacetime with bulk metric $g_{\mu\nu}$
which satisfies the {Einstein} equation {with
matter}
\begin{equation}
G_{\mu\nu}=R_{\mu\nu}-\frac{R}{2}g_{\mu\nu}=-\Lambda g_{\mu\nu}-T_{\mu\nu},\qquad (\mu,\nu=0,...p+1).
\end{equation}
Here $\Lambda$ is {a} cosmological constant, {and
$T_{\mu\nu}$ is the energy-momentum tensor of matter fields.}

{Now we embed} a timelike hypersurface $\Sigma_{c}$ with
induced metric $h_{ab}$ {into} this spacetime. The momentum
constraints on $\Sigma_{c}$ then could be written as
\begin{equation}
D_{a}K_{b}^{a}-D_{b}K=-T_{\mu b}n^{\mu},\ \ \ \ (a,b=0,...p),
\end{equation}
where $K_{ab}$ is the extrinsic curvature of $\Sigma_{c}$ and the
covariant derivative operator $D$ is compatible with the induced
metric, namely $D_{a}h_{bc}=0$. Besides, the Hamiltonian constraint
on $\Sigma_{c}$ has the form
\[
^{p+1}R+K^{\mu\nu}K_{\mu\nu}-K^{2}=2\Lambda+2T_{\mu\nu}n^{\mu}n^{\nu}.
\]

These {constraints} give $p+2$ equations of $K_{ab}$.
{Next} we decompose the $p+2$ dimensional Weyl tensor in
terms of the intrinsic curvature and extrinsic curvature of the
hypersurface.% since we want to derive a theory living on the
%boundary from the bulk.
It turns out that the results are
\begin{eqnarray}
C_{\mu\nu\sigma\rho}& = &R_{\mu\nu\sigma\rho}-\frac{4\left(\Lambda+T\right)}{p(p+1)}g_{\mu[\sigma}g_{\rho]\nu}
+\frac{2}{p}(g_{\mu[\sigma}T_{\rho]\nu}-g_{\nu[\sigma}T_{\rho]\mu}),
\\
C_{abcd} & = & ^{p+1}R_{abcd}-K_{ac}K_{bd}+K_{ad}K_{bc}-\frac{4(\Lambda+T)}{p(p+1)}h_{a[c}h_{d]b}\nonumber \\
 &  & +\frac{2}{p}h_{a}^{\alpha}h_{b}^{\beta}h_{c}^{\gamma}h_{d}^{\delta}(g_{\alpha[\gamma}T_{\delta]\beta}
 -g_{\beta[\gamma}T_{\delta]\alpha}),\label{eq:5}
\\
C_{abc(n)}& = &D_{a}K_{bc}-D_{b}K_{ac}+\frac{2}{p}h_{a}^{\alpha}h_{b}^{\beta}h_{c}^{\gamma}
(g_{\alpha[\gamma}T_{\delta]\beta}-g_{\beta[\gamma}T_{\delta]\alpha})n^{\delta},\label{eq:6}
\\
C_{a(n)b(n)} & = & -\ ^{p+1}R_{ab}+KK_{ab}-K_{ac}K_{b}^{c}+h_{a}^{\alpha}h_{b}^{\gamma}R_{\alpha\gamma}
-\frac{2\left(\Lambda+T\right)}{p(p+1)}h_{ab}\nonumber \\
 &  & +\frac{2}{p}h_{a}^{\alpha}h_{b}^{\gamma}(g_{\alpha[\gamma}T_{\delta]\beta}
 -g_{\beta[\gamma}T_{\delta]\alpha})n^{\beta}n^{\delta}.\label{eq:7}
\end{eqnarray}
Here $n^{\mu}$ is the unit normal vector of $\Sigma_{c}$ and $C_{abc(n)}=C_{abc\mu}n^{\mu},C_{a(n)b(n)}=C_{a\mu b\nu}n^{\mu}n^{\nu},T=T_{\ \mu}^{\mu}$.

{T}he Petrov-like conditions is defined as
\begin{equation}
C_{(l)i(l)j}=l^{\mu}m_{i}^{\nu}l_{\sigma}m_{j}^{\rho}C_{\mu\nu\sigma\rho}=0\label{eq:8}
\end{equation}
{on the hypersurface $\Sigma_{c}$.} Here the $p+2$ Newman-Penrose-like vector fields should satisfy the
relations
\begin{equation}
l^{2}=k^{2}=0,\text{\ \ }(k,l)=1,\ \ (k,m_{i})=(l,m_{i})=0,\ \ (m^{i},m_{j})=\delta_{j}^{i}.
\end{equation}

{Furthermore, if we introduce the Brown-York tensor on the
hypersurface which is defined as}
\begin{equation}
t_{ab}=Kh_{ab}-K_{ab}\label{eq:10},
\end{equation}
then with equations (\ref{eq:5})-(\ref{eq:7}), the Petrov-like
conditions (\ref{eq:8}) {can be expressed in terms of the
Brown-York tensor. When the induce metric and the intrinsic
curvature are fixed for the cutoff surface, we find that the only
dynamical variables for gravity {are} the Brown-York tensor.}

{In the absence of} matter field{s}, the Petrov-like condition on $\Sigma_{c}$ gives $\frac{p(p+1)}{2}-1$ equations
of the Weyl tensor. It reduces the $\frac{(p+2)(p+1)}{2}$
components of the extrinsic curvature $K_{ab}$ to $p+2$
unconstrained variables, and so is the Brown-York tensor, which
may be interpreted as the energy density, velocity field $v^{i}$
and pressure $P$ of a fluid living on the hypersurface. The $p+2$
momentum and Hamiltonian constraints on $\Sigma_{c}$ then become
an equation of state and evolution for the fluid variables.
However, without any further expansion in the fluid described here
has rather exotic dynamical equations\cite{key-18}. In the
following section, we will do this expansion around a large mean
curvature limit and get the familiar hydrodynamical equation
{in the near horizon limit}.

{In the presence of} matter field{s}, {obviously} we need {further
impose appropriate boundary conditions for matter field to constrain
the degrees of freedom of matter on the surface such that the total
remaining degrees of freedom is what we need for a fluid living on
the surface. In general, such boundary conditions depends on the
content of matter fields as well as the property of the dual fluid.
{Obviously, a better understanding for Strominger's boundary
condition should be of benefit to find boundary condition for matter
field. Now let's reconsider Strominger's Petrov like boundary
condition. For the gravity/fliud correspondence, the bulk is a p+2
dim manifold which contains horizon and a time-like
boundary\cite{key-13}. For vacuum case, such system can be
controlled by the initial-boundary value problem(IBVP)
method\cite{FN99}. Based on the work by Friedrich and Nagy, it is
easy to see that Strominger's boundary condition is closed related
with free boundary data of IBVP. In other words, we can say
Strominger's boundary condition is coming from the free boundary
data of IBVP of vacuum Einstein system. This observation gives us a
guideline for searching a suitable boundary condition for matter
field.} In this paper we intend to take the electromagnetic field as
an example and propose a boundary condition which is very analogous
to the Petrov-like condition for gravity. Specifically, the
energy-momentum tensor of the electromagnetic field is given by}
\begin{eqnarray}
T_{\mu\nu} & = &
\frac{1}{4}g_{\mu\nu}F_{\rho\sigma}F^{\rho\sigma}-F_{\mu\rho}F_{\nu}^{\;\rho}.
\end{eqnarray}

{Based on the result of IBVP of Maxwell field\cite{FN99} }, we
{propose} the boundary condition {for the Maxwell field to be}
\begin{equation}
F_{(l)i}=F_{\mu\text{\ensuremath{\nu}}}l^{\mu}m_{i}^{\nu}=0\label{eq:11}.
\end{equation}
{Physically,
such a boundary condition can be intuitively
understood as there is no outgoing electromagnetic waves passing through
the cutoff surface.} Similarly as we fix the induced metric $h_{ab}$ on the cutoff
surface, we also fix {$F_{ab}|_{\Sigma_{c}}$,}
which could be regarded as the Dirichlet-like boundary condition. Here
$a,b$ stand the components on hypersurface. Mathematically, these equations give
$p+\frac{p(p+1)}{2}$ constraints for Maxwell field and only one
component is remaining now. While we have the current conservation
law $D_{a}J^{a}=0$. With identification $J^{a}=-n_{\mu}F^{\mu a}$
on boundary\cite{key-5}, it gives
\begin{eqnarray}
-D_{a}J^{a} & = & D_{a}\left(n_{\mu}F^{\mu a}\right)=h_{\beta}^{\alpha}\nabla_{\alpha}\left(n_{\mu}F^{\mu\beta}\right)\nonumber \\
 & = & F^{\mu\beta}h_{\beta}^{\alpha}\nabla_{\alpha}n_{\mu}+n_{\mu}h_{\beta}^{\alpha}\nabla_{\alpha}F^{\mu\beta}\nonumber \\
 & = & F^{\mu\beta}K_{\mu\beta}+n_{\mu}n_{\beta}\nabla_{\alpha}F^{\mu\beta}+n_{\mu}\nabla_{\alpha}F^{\mu\alpha}=0.
\end{eqnarray}
The perturbation of this conservation law will be presented
{in} the appendix. It is clear that it gives a constraint to
the Maxwell field.

Based on the above analysis, we {find the total degrees of
freedom for gravity reduce to those for a \emph{charged} fluid.
This match makes it possible to construct a duality} between the
gravity and fluid. {To prove this, the key procedure is to
obtain} the hydrodynamical equations for a fluid from the bulk
Einstein equation. This is {what we intend to do} in the
following section{s}.

\section{From Petrov-Einstein to Navier-Stokes in charged AdS black brane}

In this section, {we investigate} the dynamic behavior of
the geometry on a {flat embedding}. {Specifically},
taking a timelike hypersurface in {a} charged $AdS_{p+2}$
black brane, we impose {boundary conditions for both
gravity and} Maxwell fields in terms of Brown-York tensor and
electromagnetic field tensor. {Then we expand the
perturbations of the extrinsic curvature as well as the Maxwell
field} around a large mean curvature, {which in our cases
indicates that the cutoff surface approaches to the horizon of the
black hole.} We find that the incompressible Navier-Stokes
equation can be derived with the help of the Einstein momentum
constraints.

\subsection{Petrov-like conditions of Charged $AdS_{p+2}$ black brane}

The spacetime we {consider} here is {a} charged
$AdS_{p+2}$ black brane. Its metric {reads as}
\begin{equation}
ds_{p+2}^{2}=-f(r)dt^{2}+2dtdr+r^{2}\delta_{ij}dx^{i}dx^{j},\ (i,j=1,...p),\nonumber
\end{equation}
\begin{equation}
f(r)=\frac{r^{2}}{l^{2}}-\frac{2\mu}{r^{p-1}}+\frac{Q^{2}}{r^{2p-2}}.\label{eq:14}
\end{equation}
Here Q is the electric charge. Take an embedd{ed}
hypersurface $\Sigma_{c}$ by setting $r=r_{c}$ such that the
induced metric can be written as
\begin{equation}
ds_{p+1}^{2}=-f(r_{c})dt^{2}+r_{c}^{2}\delta_{ij}dx^{i}dx^{j}.
\end{equation}
It {is} obvious that $\Sigma_{c}$ is an
{intrinsically} flat $p+1$ dimensional spacetime. In order to
discuss the dynamical behavior of the geometry in the
non-relativistic limit, we introduce a parameter
$\lambda$ which rescales the time coordinate by $\tau=\lambda
x^{0}=\lambda\sqrt{f_{c}}t$. The induced metric on $\Sigma_{c}$
then could be written as
\begin{eqnarray}
ds_{p+1}^{2} & =-\left(dx^{0}\right)^{2}+r_{c}^{2}\delta_{ij}dx^{i}dx^{j}\nonumber \\
 & =-\frac{1}{\lambda^{2}}d\tau^{2}+r_{c}^{2}\delta_{ij}dx^{i}dx^{j}.\label{eq:13}
\end{eqnarray}
Moreover, we identify $r_{c}-r_{h}=\alpha^{2}\lambda^{2}$ so that
the limit $\lambda\rightarrow0$ can also be thought {of} as a
kind of near horizon limit\cite{key-18}. Here $r_{h}$ is the
horizon of bulk spacetime, namely, $f(r_{h})=0$. $\alpha$ is a
parameter which {will be fixed} later. It turns out that
$\alpha$ {depends} on the {specific form of the
spacetime metric}.

Then we turn to the Petrov-like condition. The base vector fields taken
here are
\begin{equation}
m_{i}=\frac{1}{r}\partial_{i},\quad\sqrt{2}l=\frac{1}{\sqrt{f}}\partial_{t}-n=
\partial_{0}-n,\quad\sqrt{2}k=-\frac{1}{\sqrt{f}}\partial_{t}-n=-\partial_{0}-n.\label{eq:17}
\end{equation}
in (\ref{eq:8}), so the Petrov-like condition becomes
\begin{equation}
C_{0i0j}+C_{0ijn}+C_{0jin}+C_{ninj}=0.\label{eq:15}
\end{equation}
{This} equation may be {expressed} in terms of the
Brown-York tensor on $\Sigma_{c}$. Its components in the coordinate
{system} (\ref{eq:13}) are
\begin{equation}
t=pK,\ \ t_{i}^{\tau}=-K_{i}^{\tau},\ \ t_{j}^{i}=K\delta_{j}^{i}-K_{j}^{i},\ \ t_{\tau}^{\tau}=K-K_{\tau}^{\tau}.\label{eq:16}
\end{equation}
In terms of the above variables, equation (\ref{eq:15})
turns out to be
\begin{eqnarray}
0 & = & \frac{2}{\lambda^{2}}t_{i}^{\tau}t_{j}^{\tau}+\frac{t^{2}}{p^{2}}h_{ij}
-\frac{t}{p}t_{\tau}^{\tau}h_{ij}+t_{\tau}^{\tau}t_{ij}+2\lambda\partial_{\tau}
(\frac{t}{p}h_{ij}-t_{ij})-\frac{2}{\lambda}D_{\text{(}i}t_{j)}^{\tau}
-t_{ik}t_{j}^{k}\nonumber \\
 &  & +\frac{1}{p}(T_{\delta\beta}n^{\beta}n^{\delta}+2\Lambda+T+T_{00}
 -2T_{\delta0}n^{\delta})g_{ij}-T_{ij}.\label{eq:20}
\end{eqnarray}

Now we need to express the energy-momentum tensor $T_{\mu\nu}$ as
the electromagnetic field tensor $F_{\mu\nu}$. This will involve the
rise of lower index of the $p+2$ dimensional variables $F_{\mu\nu}$.
Since the Maxwell field is $p+2$ dimensional, and the bulk metric is
disturbed, it should be more cautious to deal with the Maxwell field.
{To express the energy-momentum tenser, we also need the
behavior of the metric. Based on {the method in \cite{key-18}}, what we need
is only the near horizon metric because the boundary will approach
to the horizon under the large mean curvature limit. To control the
near horizon metric, Bondi-like coordinates are always a convenient
choice.} This means that we could fix
$g_{rt}=1$ and $g_{ri}=0$ on $\Sigma_{c}$, which is equivalent to
$g^{rt}=1,\ g^{ti}=0$\cite{wu08}. Here $i$ is the pure space
component index. With metric (\ref{eq:14}) and (\ref{eq:17}),
the boundary condition for electromagnetic field
could be written as
$F_{ti}|_{\Sigma_{c}}=0$ and $F_{ij}|_{\Sigma_{c}}=0$.
Together with constraints (\ref{eq:11}), it leads to $F_{ri}=0$ and
$F^{ti}=0$.
With these
constraints for Maxwell field and the specific coordinate we have chosen,
$F^{\mu\nu}$ and $F_{\mu}^{\ \nu}$ could be written as in terms of
$F_{\mu\nu}$ on $\Sigma_{c}$.
\begin{eqnarray}
F^{rt}|_{r_{c}}& = &F_{tr},\ \ F^{ri}|_{r_{c}}=F_{tr}g^{ri},\ \ F^{ij}|_{r_{c}}=0,\nonumber \\
F_{\ t}^{r}|_{r_{c}} & = & F_{rt}g^{rr},\ \ F_{\ i}^{r}|_{r_{c}} =0,\ \
F_{t}^{\ i}|_{r_{c}} =F_{tr}g^{ri}.\label{eq:21-1}
\end{eqnarray}
After some straightforward calculations (see appendix A
{for details}), the Petrov-like condition (\ref{eq:20})
can be written as
\begin{eqnarray}
0&= & \frac{2}{\lambda^{2}}t_{i}^{\tau}t_{j}^{\tau}+\frac{t^{2}}{p^{2}}h_{ij}
-\frac{t}{p}t_{\tau}^{\tau}h_{ij}+t_{\tau}^{\tau}t_{ij}
+2\lambda\partial_{\tau}(\frac{t}{p}h_{ij}-t_{ij})-\frac{2}{\lambda}
D_{\text{(}i}t_{j)}^{\tau}-t_{ik}t_{j}^{k}\nonumber \\
 & & +\frac{1}{p}\left(\lambda^{2}fF_{r\tau}F_{r\tau}+2\Lambda\right)\delta_{j}^{i}.\label{eq:19}
\end{eqnarray}
So far, all the calculations are exact. Next we
will disturb the Einstein-Maxwell field in order to get the hydrodynamic
equation.

\subsection{Perturbation of Einstein-Maxwell field}

Through this paper the induced metric of background is fixed such
that {the dynamical variable is the extrinsic curvature}.
{W}e consider {the perturbations }of the extrinsic
curvature on the {cutoff surface}. {For the charged
AdS black brane, it is straightforward to obtain} the components
of the extrinsic curvature of $\Sigma_{c}$ {as}
\begin{eqnarray}
K_{j}^{i}=\frac{\sqrt{f}}{r}\delta_{j}^{i}, & \ \ \ \ K_{\tau}^{\tau}=\frac{1}{2\sqrt{f}}\partial_{r}f,\nonumber \\
K_{i}^{\tau}=0, & \ \ \ \ K=\frac{1}{2\sqrt{f}}\partial_{r}f+\frac{p\sqrt{f}}{r}.
\end{eqnarray}
The spacetime is disturbed by adding a corresponding term which
has a higher order than the background. In terms of Brown-York tensor,
these are
\begin{eqnarray}
t_{i}^{\tau} & \to & 0+\lambda t_{i}^{\tau(1)}+\cdots,\nonumber \\
t_{\tau}^{\tau} & \to & \frac{p\sqrt{f}}{r}+\lambda t_{\tau}^{\tau(1)}+\cdots,\nonumber \\
t_{j}^{i} & \to & \left(\frac{1}{2\sqrt{f}}\partial_{r}f+\frac{(p-1)\sqrt{f}}{r}\right)\delta_{j}^{i}+\lambda t_{j}^{i(1)}+\cdots,\nonumber \\
t & \to & \frac{p}{2\sqrt{f}}\partial_{r}f+\frac{p^{2}\sqrt{f}}{r}+\lambda t^{(1)}+\cdots.\label{eq:21}
\end{eqnarray}
{All the perturbation terms} are at order $O(\lambda^{1})$.
{The next key step in our formalism is to consider the
behavior of these perturbations as the cutoff surface approaches
the black brane horizon. As mentioned above, our strategy is identifying the
perturbation parameter $\lambda$ with the location of the
hypersurface $r_{c}-r_{h}=x$ by $x=\alpha^{2}\lambda^{2}$, such
that the near horizon limit can be achieved simultaneously with
the non-relativistic limit.} As a
result, the following quantities should also be expanded as
\begin{eqnarray}
\frac{1}{r_{c}} & = & \frac{1}{r_{h}}\left(1-\frac{x}{r_{h}}\right)+\cdots=\frac{1}{r_{h}}
-\frac{1}{r_{h}^{2}}\alpha^{2}\lambda^{2}+\cdots,\nonumber \\
f{}_{c} & = & \frac{\left(r_{h}+x\right)^{2}}{l^{2}}-\frac{2\mu}{\left(r_{h}+x\right)^{p-1}}
+\frac{Q^{2}}{\left(r_{h}+x\right)^{2p-2}}\nonumber \\
 & = & bx+cx^{2}+\cdots=b\alpha^{2}\lambda^{2}+c\alpha^{4}\lambda^{4}+\cdots,\nonumber \\
h^{ij}|_{r_{c}} & = & \frac{1}{r_{h}^{2}}\left(1-\frac{2x}{r_{h}}\right)\delta^{ij}+\cdots=\left(\frac{1}
{r_{h}^{2}}-\frac{2\alpha^{2}\lambda^{2}}{r_{h}^{3}}\right)\delta^{ij}+\cdots\nonumber \\
 & = & h^{ij(0)}+h^{ij(1)}+\cdots.
\end{eqnarray}
Here the brief notes $
b=\frac{2r_{h}}{l^{2}}-\frac{2\mu\left(1-p\right)}{r_{h}^{p}}
+\frac{\left(2-2p\right)Q^{2}}{r_{h}^{2p-1}}$,
$c=\frac{1}{l^{2}}+\frac{\mu p\left(1-p\right)}{r_{h}^{p+1}}
+\frac{\left(1-p\right)\left(1-2p\right)Q^{2}}{r_{h}^{2p}}$ and
$h^{ij(0)}=\frac{1}{r_{h}^{2}}\delta^{ij}, h^{ij(1)}=\frac{2\alpha^{2}\lambda^{2}}{r_{h}^{3}}\delta^{ij}
$
have been used.

{Now we} consider the {perturbations} of the Maxwell
field. The background field of $AdS_{p+2}$ black brane is
\begin{equation}
F=\sqrt{p(p-1)}\frac{Q}{r^{p}}dt\wedge dr=Cdt\wedge dr,
\end{equation}
where Q is the {electric} charge and
$C=\sqrt{p(p-1)}\frac{Q}{r^{p}}$. The perturbation of Maxwell
field can be written as
\begin{equation}
F_{r\tau} =\frac{1}{\lambda\sqrt{f}}C+F_{r\tau}^{(0)}.\label{eq:25}
\end{equation}

There is a need to explain why we take this perturbation. The
order of $F_{r\tau}$ is $O(\lambda^{-1})$ before we identify the
non-relativistic limit and the near horizon
limit. So the perturbation has just a higher order than
the background. This is a natural way. In order to calculate the
order of the electromagnetic field with upper index, we need to
work out the order of the inverse metric. Here we point out that
under appropriate coordinates, there are $g^{rr}\sim
O(\lambda^{2}),\ g^{ri}\sim O(\lambda^{2}),\ g^{ij}\sim
O(\lambda^{0}),\ (i=1,...p)$\cite{wu08}. With these
preparations, the Petrov-like conditions (\ref{eq:19}) on $\Sigma_{c}$
turn out to be order by order
\begin{equation}
\lambda^{-2}:-\frac{b}{4\alpha^{2}}\delta_{j}^{i}+\frac{b}{4\alpha^{2}}\delta_{j}^{i}=0,
\end{equation}
\begin{equation}
\lambda^{0}:\frac{\sqrt{b}}{\alpha}t_{j}^{i(1)}= 2h^{ik(0)}t_{k}^{\tau(1)}t_{j}^{\tau(1)}-2h^{ik(0)}t_{(j,k)}^{\tau(1)}+\frac{\sqrt{b}}{\alpha}\frac{t^{(1)}}{p}\delta_{j}^{i} +\frac{1}{p}\left(C_{h}^{2}+2\Lambda\right)\delta_{j}^{i}.\label{eq:28}
\end{equation}
Here $C_{h}=\sqrt{p(p-1)}\frac{Q}{r_{h}^{p}}$ is a constant which
is relevant to the background electromagnetic field. It's obvious
that the background ($\lambda^{-2}$ order ) satisfies this condition
automatically.

\subsection{Navier-Stokes equation in charged AdS black brane}

At last we come to the momentum constraints on $\Sigma_{c}$. Since
$\Sigma_{c}$ is {intrinsically} flat, the momentum
constraints can be written as
\begin{equation}
\partial_{a}t_{b}^{a}=T_{\mu b}n^{\mu},\ (a,b=0,...p).\label{eq:30}
\end{equation}
The time component {on} the {left-hand} side could
be expanded as
\begin{equation}
\partial_{\mu}t_{\tau}^{\mu}=\partial_{\tau}t_{\tau}^{\tau}+\partial_{i}t_{\tau}^{i}.
\end{equation}
At leading order, there are
\begin{align}
\partial_{\text{\ensuremath{\tau}}}t_{\tau}^{\tau} & =\lambda\partial_{\text{\ensuremath{\tau}}}t_{\tau}^{\tau(1)}+O(\lambda^{2})=O(\lambda),\nonumber \\
\partial_{i}t_{\tau}^{i} & =\partial_{i}\left(t^{i\mu}h_{\mu\tau}\right)=-\frac{1}{\lambda^{2}}\partial_{i}
t^{i\tau}=-\frac{1}{r^{2}\lambda^{2}}\partial_{i}t_{i}^{\tau}=O(\lambda^{-1}).
\end{align}
The {time component on the right-hand side of (\ref{eq:30})} is
\begin{eqnarray}
T_{\tau}^{\mu}n_{\mu}  =  \frac{1}{\sqrt{f}}T_{\tau}^{r}=-\frac{1}{\lambda f}F^{r\rho}F_{t\rho}=0.\label{eq:31}
\end{eqnarray}
So the time component of momentum constraints at leading order turns to
\begin{equation}
\partial_{i}t^{\tau i(1)}=0.
\end{equation}
Identifying
\begin{equation}
t_{i}^{\tau(1)}=\frac{v_{i}}{2},\;\ \ \ \frac{t^{(1)}}{p}=\frac{P}{2}.\label{eq:35}
\end{equation}
We get the incompressible condition.
\begin{equation}
\partial_{i}v^{i}=0.
\end{equation}
Then consider the space components of the momentum constraints. The
left hand side could be expanded as
\begin{equation}
\partial_{\mu}t_{i}^{\mu}=\partial_{\tau}t_{i}^{\tau}+\partial_{k}t_{i}^{k},
\end{equation}
At leading order
\begin{eqnarray}
\partial_{\tau}t_{i}^{\tau} & =\lambda\partial_{\tau}t_{i}^{\tau(1)}+O(\lambda^{2}),\nonumber \\
\partial_{k}t_{i}^{k} & =\lambda\partial_{k}t_{i}^{k(1)}+O(\lambda^{2}).
\end{eqnarray}
The right hand side is
\begin{equation}
T_{i}^{\mu}n_{\mu} =\frac{1}{\sqrt{f}}T_{i}^{r}=-\frac{1}{\sqrt{f}}\left(F^{rt}F_{it}
+F^{rj}F_{ij}\right)=0.\label{eq:37}
\end{equation}
So the space components of the momentum constraints turn to
\begin{equation}
\partial_{\tau}t_{i}^{\tau(1)}+\partial_{k}t_{i}^{k(1)}= 0,
\end{equation}
Combining with the identifying equation (\ref{eq:35}) and (\ref{eq:28}),
we get
\begin{equation}
\partial_{\tau}v_{i}+\partial_{i}P+\frac{\alpha}{\sqrt{b}}\left(v^{k}\partial_{k}v_{i}
-\partial^{k}\partial_{k}v_{i}\right)=0.\label{eq:39}
\end{equation}
{So we see that $\alpha$ should be fixed to be $\sqrt{b}$, which leads to}
\begin{equation}
\partial_{\tau}v_{i}+\partial_{i}P+v^{k}\partial_{k}v_{i}-\partial^{k}\partial_{k}v_{i}=0.
\end{equation}
This is exact the incompressible Navier-Stokes equation in $p$ space
dimensions.

{In general}, the hydrodynamic equation {for a
charged fluid is}\cite{MH}
\begin{equation}
\partial_{\tau}v_{i}+\partial_{i}P+v^{k}\partial_{k}v_{i}-\partial^{k}\partial_{k}v_{i}
=F_{ia}J^{a}.\label{eq:e}
\end{equation}
Here $a$ contains $i$ and $\tau$. {In this equation} there
{is} a term caused by the external electromagnetic force.
{However, since $F_{ab}|_{r_{c}}=0$ on the cutoff surface for the charged AdS black brane, the right-hand side of the Navier-Stokes
equation vanishes.} In section V, {we will demonstrate}
that an external force does arise when the background
electromagnetic field {$F_{ab}|_{r_{c}}\ne 0$, which corresponds to hydrodynamics of a magnetofluid}.

\section{From Petrov-Einstein to Navier-Stokes in RN-AdS spacetime}

The hypersurface in charged AdS black brane is intrinsic
flat. Now let us consider an intrinsically curved hypersurface in Reissner-Nordstrom-AdS
(RN-AdS) spacetime and study the influence of the space curvature. The bulk metric is
\[
ds_{p+2}^{2}=-f(r)dt^{2}+2dtdr+r^{2}d\Omega_{p}^{2},
\]
\begin{equation}
f(r)=1-\frac{2\mu}{r^{p-1}}+\frac{Q^{2}}{r^{2p-2}}+\frac{r^{2}}{l^{2}}.\label{eq:41}
\end{equation}
Here $d\Omega_{p}^{2}$ is p-dimensional spherical metric which could
be written as
\begin{equation}
d\Omega^{2}=d\theta_{1}^{2}+sin^{2}\theta_{1}d\theta_{2}^{2}+\cdots
+sin^{2}\theta_{1}\cdots sin^{2}\theta_{p-1}d\theta_{p}^{2}.
\end{equation}
The metric of the embedding hypersurface $\Sigma_{c}$ {at}
$r=r_{c}$ is
\begin{equation}
ds_{p+1}^{2}=-f(r_{c})dt^{2}+r_{c}^{2}d\Omega_{p}^{2}.
\end{equation}
Introducing a coordinate transformation $\tau=\lambda x^{0}=\lambda\sqrt{f_{c}}t$, the induced
metric could be written as
\begin{eqnarray}
ds_{p+1}^{2} & =-\left(dx^{0}\right)^{2}+r_{c}^{2}d\Omega_{p}^{2}\nonumber \\
 & =-\frac{1}{\lambda^{2}}d\tau^{2}+r_{c}^{2}d\Omega_{p}^{2}.
\end{eqnarray}
The form of the extrinsic curvature is just the same as
{that} of charged AdS black brane. While the
Newman-Penrose-like vector fields for RN-AdS take the following
form
\begin{equation}
m_{i}=\frac{1}{rsin\theta_{1}...sin\theta_{i-1}}\partial_{i},\ \sqrt{2}l=\frac{1}{\sqrt{f}}\partial_{t}-n=\partial_{0}-n,\ \sqrt{2}k=\frac{-1}{\sqrt{f}}\partial_{t}-n=-\partial_{0}-n.
\end{equation}
The explicit expression of the Petrov-like condition is the same as (\ref{eq:15}).

Introducing Brown-York tensor $t_{ab}=Kh_{ab}-K_{ab}$, the Petrov-like
condition in intrinsically curved spacetime with matter field could be written as.
\begin{eqnarray}
0 & = & \frac{2}{\lambda^{2}}t_{i}^{\tau}t_{j}^{\tau}+\frac{t^{2}}{p^{2}}h_{ij}-\frac{t}{p}t_{\tau}^{\tau}h_{ij}+t_{\tau}^{\tau}t_{ij}+2\lambda\partial_{\tau}(\frac{t}{p}h_{ij}-t_{ij})-\frac{2}{\lambda}D_{\text{(}i}t_{j)}^{\tau}-t_{ik}t_{j}^{k}\nonumber \\
 &  & +\frac{1}{p}(T_{\delta\beta}n^{\beta}n^{\delta}+2\Lambda+T+T_{00}-2T_{\delta0}n^{\delta})g_{ij}-T_{ij}-\ ^{p+1}R_{ij}.\label{eq:47}
\end{eqnarray}
Here $^{p+1}R_{0i0j}=0$ has been considered.

Now it is needed to express $T_{\delta\beta}$ in terms of the electromagnetic
field $F_{\delta\beta}$. As have done after (\ref{eq:20}), we choose
the same coordinate and constraints for electromagnetic field.
After some similar calculations, (\ref{eq:47}) turns to
\begin{eqnarray}
0&= & \frac{2}{\lambda^{2}}t_{i}^{\tau}t_{j}^{\tau}+\frac{t^{2}}{p^{2}}h_{ij}-\frac{t}{p}t_{\tau}^{\tau}h_{ij}+t_{\tau}^{\tau}t_{ij}+2\lambda\partial_{\tau}(\frac{t}{p}h_{ij}-t_{ij})-\frac{2}{\lambda}D_{\text{(}i}t_{j)}^{\tau}-t_{ik}t_{j}^{k}\nonumber \\
 & & +\frac{1}{p}\left(\lambda^{2}fF_{r\tau}F_{r\tau}+2\Lambda\right)\delta_{j}^{i}-{}^{p+1}R_{ij}.
\end{eqnarray}
Here $^{p+1}R_{ij}$ is only relevant to the pure space components
$\theta_{i}$ and has the order $O(\lambda^{0})$.

{Now we consider} the perturbations of the Einstein-Maxwell
field. Let {us} think about gravity firstly. Taking
Brown-York tensor as the fundamental variables, the perturbation
of gravity has just the form of (\ref{eq:21}).

Identifying $r_{c}-r_{h}=x$ with $\alpha^{2}\lambda^{2}$, we get
the following expansion
\begin{eqnarray}
f(r_{c}) & = & 1-\frac{2\mu}{\left(r_{h}+x\right)^{p-1}}+\frac{Q^{2}}{\left(r_{h}+x\right)^{2p-2}}+\frac{\left(r_{h}+x\right)^{2}}{l^{2}}\nonumber \\
 & = & bx+cx^{2}+\cdots=b\alpha^{2}\lambda^{2}+c\alpha^{4}\lambda^{4}+\cdots.
\end{eqnarray}
Here
$b=-\frac{2\mu(1-p)}{r_{h}^{p}}+\frac{Q^{2}(2-2p)}{r_{h}^{2p-1}}+\frac{2r_{h}}{l^{2}}$ and
$c =\frac{\mu p(1-p)}{r_{h}^{p}}+\frac{Q^{2}(1-p)(1-2p)}{r_{h}^{2p-1}}+\frac{1}{l^{2}}$.

The perturbation of electromagnetic field takes the same form as equation
(\ref{eq:25}). At the first non-trivial order, it gives
\begin{equation}
\lambda^{0}:\ \frac{\sqrt{b}}{\alpha}t_{j}^{i(1)}= 2h^{ik(0)}t_{k}^{\tau(1)}t_{j}^{\tau(1)}-2h^{ik(0)}t_{(j,k)}^{\tau(1)}
+\frac{\sqrt{b}}{\alpha}\frac{t^{(1)}}{p}\delta_{j}^{i} +\frac{1}{p}\left(C_{h}^{2}+2\Lambda\right)\delta_{j}^{i}-{}^{p+1}R_{j}^{i}.\label{eq:52}
\end{equation}
Here $C_{h}=\sqrt{p(p-1)}\frac{Q}{r_{h}^{p}}$ is a constant which
is relevant to the background electromagnetic field.

Now considering the momentum constraints on $\Sigma_{c}$,
\begin{equation}
D_{a}t_{b}^{a}=T_{\mu b}n^{\mu},\ (a,b=0,...p).
\end{equation}
The time component gives at leading order
\begin{equation}
D_{i}t^{\tau i(1)}=0.
\end{equation}
Identifying
\begin{equation}
t_{i}^{\tau(1)}=\frac{v_{i}}{2},\;\ \ \ \frac{t^{(1)}}{p}=\frac{P}{2}.
\end{equation}
the incompressible equation is derived
\begin{equation}
D_{i}v^{i}=0.\label{eq:59}
\end{equation}
The space components of the right hand side turn out to be zero which
is similar to (\ref{eq:37}). With equations (\ref{eq:52})(\ref{eq:59})
we get
\begin{equation}
\partial_{\tau}v_{i}+D_{i}P+\frac{\sqrt{b}}{\alpha}\left(v^{j}D_{j}v_{i}
-D^{j}D_{j}v_{i}-R_{i}^{j}v_{j}\right)=0.
\end{equation}
from the momentum constraints. Here $R_{i}^{j}\propto\gamma_{i}^{j}$
has been considered since the pure space has a spherical metric. {Again} let $\alpha=\sqrt{b}$,
we get the standard incompressible Navier-Stokes equation in spatially
curved spacetime.
\begin{equation}
\partial_{\tau}v_{i}+D_{i}P+v^{j}D_{j}v_{i}-D^{j}D_{j}v_{i}-R_{i}^{j}v_{j}= 0.
\end{equation}

Compare to the situation of intrinsic flat hypersurface, the
hydrodynamic equation in intrinsic curved hypersurface has an
{extra} term which is relevant to the Ricci tensor of the
boundary. Similar to the case of charged AdS black brane, the
electromagnetic has no influence on the hydrodynamic equation.
This is a natural result because of the boundary conditions we have
taken. $F_{ab}|_{r_{c}}=0$ isolate the influence of the
electromagnetic field and lead to the incompressible Navier-Stokes
equation.

\section{The situation of magnetic Reissner-Nordstrom Black Hole}

In previous {sections}, the background electromagnetic
field on the {cutoff surface} {vanishes}. This leads
to the incompressible Navier-Stokes equation in the first
non-trivial leading order. Now we consider a 4-dimensional
magnetic black hole with metric
\begin{equation}
ds_{p+2}^{2}=-f(r)dt^{2}+2dtdr+r^{2}d\Omega_{p}^{2},\nonumber
\end{equation}
\begin{equation}
f(r)=1-\frac{2\mu}{r^{p-1}}+\frac{P^{2}}{r^{2}}.
\end{equation}
Here $P$ is the magnetic charge and the electromagnetic potential
is
\begin{equation}
A=P\cos d\phi.
\end{equation}
So the electromagnetic field tensor $F=-P\sin\theta d\theta\wedge d\phi$.
The space component is no longer zero. It is expected that this term
will lead to the external electromagnetic force in the Navier-Stokes
equation at last. $F^{\mu\nu}$ and $F_{\ \nu}^{\mu}$ could be written
in terms of $F_{\mu\nu}$ on $\Sigma_{c}$ as
\begin{eqnarray}
F^{rt}|_{r_{c}}=F_{tr}, & \ \ \ F^{ri}|_{r_{c}}=F_{tr}g^{ri}+F_{kj}g^{ji}g^{kr},\ \ \  & F^{ij}|_{r_{c}}=F_{lk}g^{kj}g^{li},\nonumber \\
F_{\ t}^{r}|_{r_{c}}=F_{rt}g^{rr}, & F_{\ i}^{r}|_{r_{c}}=F_{ki}g^{kr}, & F_{t}^{\ i}|_{r_{c}}=F_{tr}g^{ri}.
\end{eqnarray}
With the method in section IV, the Petrov condition takes the form
\begin{eqnarray}
0&= & \frac{2}{\lambda^{2}}t_{i}^{\tau}t_{j}^{\tau}+\frac{t^{2}}{p^{2}}h_{ij}
-\frac{t}{p}t_{\tau}^{\tau}h_{ij}+t_{\tau}^{\tau}t_{ij}
+2\lambda\partial_{\tau}(\frac{t}{p}h_{ij}-t_{ij})
-\frac{2}{\lambda}D_{\text{(}i}t_{j)}^{\tau}-t_{ik}t_{j}^{k}-\ ^{p+1}R_{ij}\nonumber \\
 & & +\frac{1}{p}\left(\frac{\lambda}{\sqrt{f}}F_{r\tau}F_{ki}g^{kr}g^{ri}
 -\frac{1}{f}F_{lj}F_{ki}g^{kr}g^{lr}g^{ji}+\lambda^{2}fF_{r\tau}F_{r\tau}
 -\frac{1}{2}F_{ij}F_{lk}g^{il}g^{jk}+2\Lambda\right)\delta_{j}^{i}\nonumber \\
 & & +F_{lm}F_{jk}g^{il}g^{mk}.
\end{eqnarray}
Next we should think about the perturbation of Einstein-Maxwell field.
The perturbation of gravity has just the form of (\ref{eq:21}),
but the perturbation of Maxwell field now should take the following
form
\begin{equation}
F_{r\tau}=F_{r\tau}^{(0)}+O\left(\lambda\right)
\end{equation}
since the background value of $F_{rt}$ is zero. Similar to the situation of RN black hole, we have the expansion near
horizon
\begin{eqnarray}
f(r_{c}) & = & 1-\frac{2\mu}{r_{h}+x}+\frac{P^{2}}{\left(r_{h}+x\right)^{2}}\nonumber\\
 & = & bx+cx^{2}+\cdots=b\alpha^{2}\lambda^{2}+c\alpha^{4}\lambda^{4}+\cdots,
\end{eqnarray}
where $b=\frac{2\mu}{r_{h}}-\frac{2P^{2}}{r_{h}^{3}}$ and $c=\frac{-2\mu}{r_{h}^{2}}+\frac{3P^{2}}{r_{h}^{3}}$.
At the first non-trivial leading order $O\left(\lambda^{0}\right)$,
the Petrov-like condition gives
\begin{eqnarray}
\frac{\sqrt{b}}{\alpha}t_{j}^{i(1)} & = & 2h^{ik(0)}t_{k}^{\tau(1)}t_{j}^{\tau(1)}-2h^{ik(0)}t_{(j,k)}^{\tau(1)}
+\frac{\sqrt{b}}{\alpha}\frac{t^{(1)}}{p}\delta_{j}^{i}-\ ^{p+1}R_{j}^{i}\nonumber\\
 &  & +\frac{1}{p}\left(\frac{1}{2}F_{ij}F_{lk}g^{il(0)}g^{jk(0)}
 +2\Lambda\right)\delta_{j}^{i}+F_{lm}F_{jk}g^{li(0)}g^{mk(0)}.\label{eq:68}
\end{eqnarray}
Now we come to the momentum constraints. It is easy to show that the
time component still gives the incompressible equation $D_{i}v^{i}=0$
under identification $(\ref{eq:35})$.

However, the space components of the momentum constraints turn out to
be different to the previous results. The left hand side of the momentum constraints
still gives the form the same as previous section. The right hand {side} gives
\begin{equation}
T_{i}^{\mu}n_{\mu}=\frac{1}{\sqrt{f}}T_{i}^{r}=-\frac{1}{\sqrt{f}}F^{rj}F_{ij}.
\end{equation}
Since $F^{ri}|_{r_{c}}=\lambda\sqrt{f}F_{\tau r}g^{ri}+F_{kj}g^{ji}g^{kr}\sim O\left(\lambda^{2}\right)$,
the order of the right hand is $O\left(\lambda\right)$ in the near
horizon limit. With identification $-n_{\mu}F^{\mu a}=J^{a}$,
we get the external electromagnetic force term $F_{ij}J^{j}$. {Again} let
$\alpha=\sqrt{b}$, the first order of the space components turns to
be the standard incompressible {magnetofluid} equation
\begin{equation}
\partial_{\tau}v_{i}+D_{i}P+v^{j}D_{j}v_{i}-D^{j}D_{j}v_{i}-R_{i}^{j}v_{j}=F_{ij}J^{j}.
\end{equation}
This is the expected form as equation (\ref{eq:e}), {where the external force term
comes from the background electromagnetic field on the cutoff surface}.

\section{Summary and Discussions}

In this paper we have generalized the framework presented in
\cite{key-18,key-19,key-20} to a spacetime with matter field. We
have demonstrated that with the help of Petrov-like condition and
Einstein-Maxwell constraints, the incompressible Navier-Stokes
equation {can} be derived {for a charged fluid living on the cutoff
surface which is embedded into a} charged AdS black brane, RN-AdS
{black hole} and magnetic black hole {respectively}. {During the
derivation imposing appropriate boundary condition for the Maxwell
field is crucial.} {In order to find a suitable boundary
condition, we use the result of initial boundary value problem of
Einstein system and Maxwell system. This also give us a guideline
for searching boundary condition of general matter field. This also
give us another way to understand the gravity/fluid correspondence
in terms of the evolution of partial differential equations.} Since
the background electromagnetic field on the cutoff surface {is
fixed} , the matter field does not {generate extra contributions to}
the incompressible Navier-Stokes equation in the charged AdS black
brane and RN-AdS spacetime due to the fact that $F_{ab}|_{r_{c}}=0$.
However, {for a black hole with magnetic charge,} the background
electromagnetic field {on the cutoff surface} is not zero, {its
coupling with charges of the fluid contributes an} external force
term {to} the hydrodynamic equation.

Now the holographic character of Petrov-like condition has been
further disclosed in the spacetime with electromagnetic
{field. We conjecture }that {it }should be a
universal method to reduce the Einstein equation to the
Navier-Stokes equation for a general spacetime in the presence of
a {horizon.}

In this paper, the correspondence between gravity and fluid is
feasible at the near horizon limit, while with the method of
directly disturbing the bulk metric, the Navier-Stokes equation
can be derived on arbitrary finite cutoff surface. So we wonder
whether the method in this paper can be applied to the non-near
horizon limit. This might be taken as further investigation.

\begin{acknowledgments}
{YL was partly supported by NSFC (10875057, 11178002), Fok
Ying Tung Education Foundation (No.111008), the key project of
Chinese Ministry of Education (No.208072), Jiangxi young
scientists (JingGang Star) program and 555 talent project of
Jiangxi Province. {YT and XW was partly supported by NSFC (11075206, 11175245).}}
\end{acknowledgments}

\section*{Appendix A: Calculations of the energy-momentum tensor of electromagnetic
field}

Now we will show the concrete details that how to get equation (\ref{eq:19}) from (\ref{eq:20}).
By definition
\[
T_{\mu\nu}=\frac{1}{4}g_{\mu\nu}F_{\rho\sigma}F^{\rho\sigma}-F_{\mu\rho}F_{\nu}^{\;\rho},
\]
we have
\begin{align*}
T_{\mu\nu}n^{\mu}n^{\nu} & =\frac{1}{4}F_{\rho\sigma}F^{\rho\sigma}-F_{\ \rho}^{\mu}F^{\nu\rho}n_{\mu}n_{\nu}=\frac{1}{4}F_{\rho\sigma}F^{\rho\sigma}
-\frac{1}{f}F_{\ \rho}^{r}F^{r\rho},\\
T & =\frac{p-2}{4}F_{\rho\sigma}F^{\rho\sigma},\\
T_{00} & =\frac{1}{f}T_{tt}=-\frac{1}{4}F_{\rho\sigma}F^{\rho\sigma}
-\frac{1}{f}F_{t\rho}F_{t}^{\ \rho},\\
T_{\delta0}n^{\delta} & =T_{0}^{\delta}n_{\delta}=T_{0}^{r}n_{r}=\frac{1}{f}T_{t}^{r}
=-\frac{1}{f}F^{r\rho}F_{t\rho},\\
T_{j}^{i} & =\frac{1}{4}\delta_{j}^{i}F_{\rho\sigma}F^{\rho\sigma}-F^{i\rho}F_{j\rho}.
\end{align*}
The second line of equation (\ref{eq:20}) then turns to
\begin{align*}
 & \frac{1}{p}(T_{\delta\beta}n^{\beta}n^{\delta}+2\Lambda+T+T_{00}-2T_{\delta0}n^{\delta})\delta_{j}^{i}-T_{j}^{i}\\
= & \frac{1}{p}(-\frac{1}{2}F_{\rho\sigma}F^{\rho\sigma}-\frac{1}{f}F_{\ \rho}^{r}F^{r\rho}+2\Lambda-\frac{1}{f}F_{t\rho}F_{t}^{\ \rho}+\frac{2}{f}F^{r\rho}F_{t\rho})\delta_{j}^{i}+F^{i\rho}F_{j\rho}.
\end{align*}
With equation (\ref{eq:21-1}), it is easy to work out
\begin{align*}
-\frac{1}{2}F_{\rho\sigma}F^{\rho\sigma} & =-\frac{1}{2}\left(F_{rt}F^{rt}+F_{tr}F^{tr}+F_{ij}F^{ij}\right)=F_{rt}F_{rt},\\
-\frac{1}{f}F_{\ \rho}^{r}F^{r\rho}& = -\frac{1}{f}\left(F_{\ t}^{r}F^{rt}+F_{\ i}^{r}F^{ri}\right)=\frac{1}{f}F_{rt}F_{rt}g^{rr},\\
-\frac{1}{f}F_{t\rho}F_{t}^{\ \rho} & =-\frac{1}{f}\left(F_{tr}F_{t}^{\ r}+F_{ti}F_{t}^{\ i}\right)=-\frac{1}{f}F_{tr}F_{tr}g^{rr},\\
\frac{2}{f}F^{r\rho}F_{t\rho}& =\frac{2}{f}F^{ri}F_{ti}=0,\\
F^{i\rho}F_{j\rho}& =F^{ik}F_{jk}=0.
\end{align*}
So we have
\begin{align*}
 & \frac{1}{p}(T_{\delta\beta}n^{\beta}n^{\delta}+2\Lambda+T+T_{00}-2T_{\delta0}n^{\delta})\delta_{j}^{i}-T_{j}^{i}\\
= & \frac{1}{p}[2\Lambda+F_{rt}F_{rt}]\delta_{j}^{i}\\
= & \frac{1}{p}[\lambda^{2}fF_{r\tau}F_{r\tau}+2\Lambda]\delta_{j}^{i}.
\end{align*}
Considering the perturbations of electromagnetic field
\begin{align*}
F_{r\tau} & =\frac{1}{\lambda\sqrt{f}}C_{h}+F_{r\tau}^{(0)}.
\end{align*}
We finally get
\begin{align*}
 & \frac{1}{p}(T_{\delta\beta}n^{\beta}n^{\delta}+2\Lambda+T+T_{00}-2T_{\delta0}n^{\delta})\delta_{j}^{i}-T_{j}^{i}\\
= & \frac{1}{p}(C_{h}^{2}+2\Lambda)\delta_{j}^{i}+O(\lambda).
\end{align*}
The situation of RN black hole and magnetic black hole are similar.

\section*{Appendix B: Hamiltonian constraint of charged AdS black brane and
RN-AdS spacetime}

The Hamiltonian constraint is
\[
^{p+1}R+K^{\mu\nu}K_{\mu\nu}-K^{2}=2\Lambda+2T_{\mu\nu}n^{\mu}n^{\nu}.
\]
In terms of $t_{ab}=Kh_{ab}-K_{ab}$, it turns to
\[
^{p+1}R+\left(t_{\tau}^{\tau}\right)^{2}-\frac{2h^{ij}}{\lambda^{2}}
t_{i}^{\tau}t_{j}^{\tau}-\frac{t^{2}}{p}+t_{j}^{i}t_{i}^{j}=2\Lambda
+2T_{\mu\nu}n^{\mu}n^{\nu}.
\]
The second term on the right hand side
\begin{eqnarray*}
2T_{\mu\nu}n^{\mu}n^{\nu} & = & \frac{1}{2}F_{\rho\sigma}F^{\rho\sigma}-\frac{2}{f}F_{\ \rho}^{r}F^{r\rho}\\
 & = & \frac{2}{f}F_{rt}F_{rt}g^{rr}-F_{rt}F_{rt}\\
 & = & \frac{2}{f}\left(\lambda^{2}fF_{r\tau}F_{r\tau}g^{rr}\right)-\lambda^{2}fF_{r\tau}F_{r\tau}\\
 & = & (\frac{2}{f}g^{rr}-1)C_{h}^{2}+O(\lambda)\\
 & = & C_{h}^{2}+O(\lambda).
\end{eqnarray*}

Hamiltonian constraint at order $\lambda^{-2}$ is automatically satisfied:
\[
\frac{p}{4\lambda^{2}}-\frac{p}{4\lambda^{2}}=0.
\]
In charged AdS black brane, the subheading order $\lambda^{0}$ gives
\[
-t_{\tau}^{\tau(1)}-2h^{ij}t_{i}^{\tau(1)}t_{j}^{\tau(1)}-\frac{pb}{r_{h}}=2\Lambda+C_{h}^{2}.
\]
In RN-AdS, the subleading order $\lambda^{0}$ gives
\[
^{p+1}R-t_{\tau}^{\tau(1)}-2h^{ij}t_{i}^{\tau(1)}t_{j}^{\tau(1)}-\frac{pb}{r_{h}}=2\Lambda+C_{h}^{2}.
\]
Here $^{p+1}R$ is the scalar curvature of the hypersurface.

\section*{Appendix C: Current conservation law}

\[D_{a}J^{a}=0\Longleftrightarrow\partial_{a}\left(\sqrt{-h}J^a\right)=0.\]

Since $g^{ri}~\sim O(\lambda^{2})$, and
\begin{eqnarray*}
J^{\tau} & =& n_{r}F^{r\tau}=\lambda^{2}\sqrt{f}F_{\tau r}=\lambda C+\lambda^{2}\sqrt{f}F_{\tau r}^{(0)},\\
J^i & =&  n_{r}F^{ri}=\lambda\sqrt{f}g^{ri}F_{\tau r}=g^{ri}C+\lambda\sqrt{f}g^{ri}F_{\tau r}^{(0)}.
\end{eqnarray*}
The first non-trivial leading order gives
\[\partial_{\tau}F_{\tau r}^{(0)}=0.\]

\end{document}